\documentclass[modern]{aastex631}

\usepackage{amsmath}

\begin{document}

\title{Diurnal ejection of boulder clusters on comet 67P lasting beyond 3 AU}

\correspondingauthor{Xian Shi}
\email{shi@shao.ac.cn}

\author[0000-0002-4120-7361]{Xian Shi}
\affiliation{Shanghai Astronomical Observatory, Chinese Academy of Sciences \\
80 Nandan Road, 200030 Shanghai, China}
\affiliation{Max-Planck-Institut f{\"u}r Sonnensystemforschung \\
Justus-von-Liebig-Weg 3, 37077 G{\"o}ttingen, Germany}

\author[0000-0002-4302-5633]{Xuanyu Hu}
\affiliation{Institut f{\"u}r Geod{\"a}sie und Geoinformationstechnik, Technische Universit{\"a}t Berlin \\
Kaiserin-Augusta-Allee 104-106, 10553 Berlin, Germany.}

\author[0000-0001-6608-1489]{Jessica Agarwal}
\affiliation{Institut f{\"u}r Geophysik und Extraterrestrische Physik, Technische Universit{\"a}t Braunschweig \\
Mendelssohnstrasse 3, 38106 Braunschweig, Germany}
\affiliation{Max-Planck-Institut f{\"u}r Sonnensystemforschung \\
Justus-von-Liebig-Weg 3, 37077 G{\"o}ttingen, Germany}

\author[0000-0003-4277-1738]{Carsten G{\"u}ttler}
\affiliation{Institut f{\"u}r Planetologie, Universit{\"a}t M{\"u}nster \\
Wilhelm-Klemm-Str. 10, 48149 M{\"u}nster, Germany}
\affiliation{Max-Planck-Institut f{\"u}r Sonnensystemforschung \\
Justus-von-Liebig-Weg 3, 37077 G{\"o}ttingen, Germany}

\author{Martin Rose}
\affiliation{Ingenieurbuero Dr.-Ing. Martin Rose \\
Sommerhofenstrasse 148, 71067 Sindelfingen, Germany}

\author{Horst Uwe Keller}
\affiliation{DLR Institut f{\"u}r Planetenforschung \\
Rutherfordstraße 2, 12489 Berlin, Germany}

\author[0000-0001-8435-5287]{Marco Fulle}
\affiliation{INAF - Osservatorio Astronomico \\
Via Tiepolo 11, 34014 Trieste, Italy}

\author{Jakob Deller}
\affiliation{Max-Planck-Institut f{\"u}r Sonnensystemforschung \\
Justus-von-Liebig-Weg 3, 37077 G{\"o}ttingen, Germany}

\author{Holger Sierks}
\affiliation{Max-Planck-Institut f{\"u}r Sonnensystemforschung \\
Justus-von-Liebig-Weg 3, 37077 G{\"o}ttingen, Germany}

\received{November 15, 2023}
\revised{December 19, 2023}
\accepted{December 27, 2023}

\begin{abstract}
Ejection of large boulder-like debris is a vigorous form of cometary activity that is unlikely induced by water ice out-gassing alone but rather associated with the sublimation of super-volatile ices. Though perceived on several comets, actual pattern and mechanism of such activity are still unclear. Here we report on a specialized observation of ejections of decimeter- to meter-sized boulders on comet 67P/Churyumov-Gerasimenko outbound between 2.5 and 3.3 AU from the Sun. With a common source region, these events recurred in local morning. The boulders of elongated shapes were ejected in clusters at low inclinations comparable to the solar elevation below 40 degrees at the time. We show that these chunks could be propelled by the surrounding, asymmetric gas field that produced a distinct lateral acceleration. Possibly both water and carbon dioxide have contributed to their mobilization, while the season and local topography are among deciding factors. The mechanisms for sustaining regular activity of comets at large heliocentric distances are likely more diverse and intricate than previously thought.
\end{abstract}

\keywords{Comets (280) --- Coma dust (2159) --- Comet surfaces (2161) --- Comet volatiles (2162)}

\section{Introduction} \label{sec:intro}

Mass ejections have been discovered on ever more small bodies in our Solar System, yet the underlying mechanisms are diverse and, in many cases, unsettled \citep{2006Sci...312..561H,2010AJ....140.1519J,2019Sci...366.3544L}. One typical such phenomenon, the ejection of large boulder-like debris has been observed on several comets and is thought to be associated with the sublimation of highly volatile ices, such as $\text{CO}_2$, more potent than water activity \citep{2009Icar..203..571R,2013Icar..222..634K,2016MNRAS.462S..78A}. The European Space Agency’s Rosetta mission is the only space mission to have followed a comet through its perihelion passage. During the over-two-year rendezvous from ~4 AU inbound to ~3.8 AU outbound, Rosetta witnessed, at unprecedented spatial and temporal resolution, diversity of activity taking place on its target comet 67P/Churyumov-Gerasimenko (67P). Multi-instrument investigations have substantiated the intricate nature of different types of activity. Regular activity, in the form of global emission of predominantly micrometer- to centimeter-sized dust particles from the sunlit surface areas is driven by the sublimation of water and $\text{CO}_2$ ices in the top layer of the nucleus \citep{2015Sci...347a1044S,2018NatAs...2..562S,2017MNRAS.469S.755B,2022NatAs...6..546C,2020MNRAS.493.4039F}. Various types of transient activity, such as spontaneous outbursts more frequent around perihelion, could be of different origins, including intensive sublimation of buried super-volatile ice reservoirs, landslides and cliff collapses, or crystallisation of amorphous ice \citep{2016A&A...593A..76S,2016MNRAS.462S.184V,2016MNRAS.462S.220G,2017MNRAS.469S.606A}.

\section{Observations} \label{sec:data} 

\begin{figure}[ht]
\centering
\includegraphics[width=\textwidth]{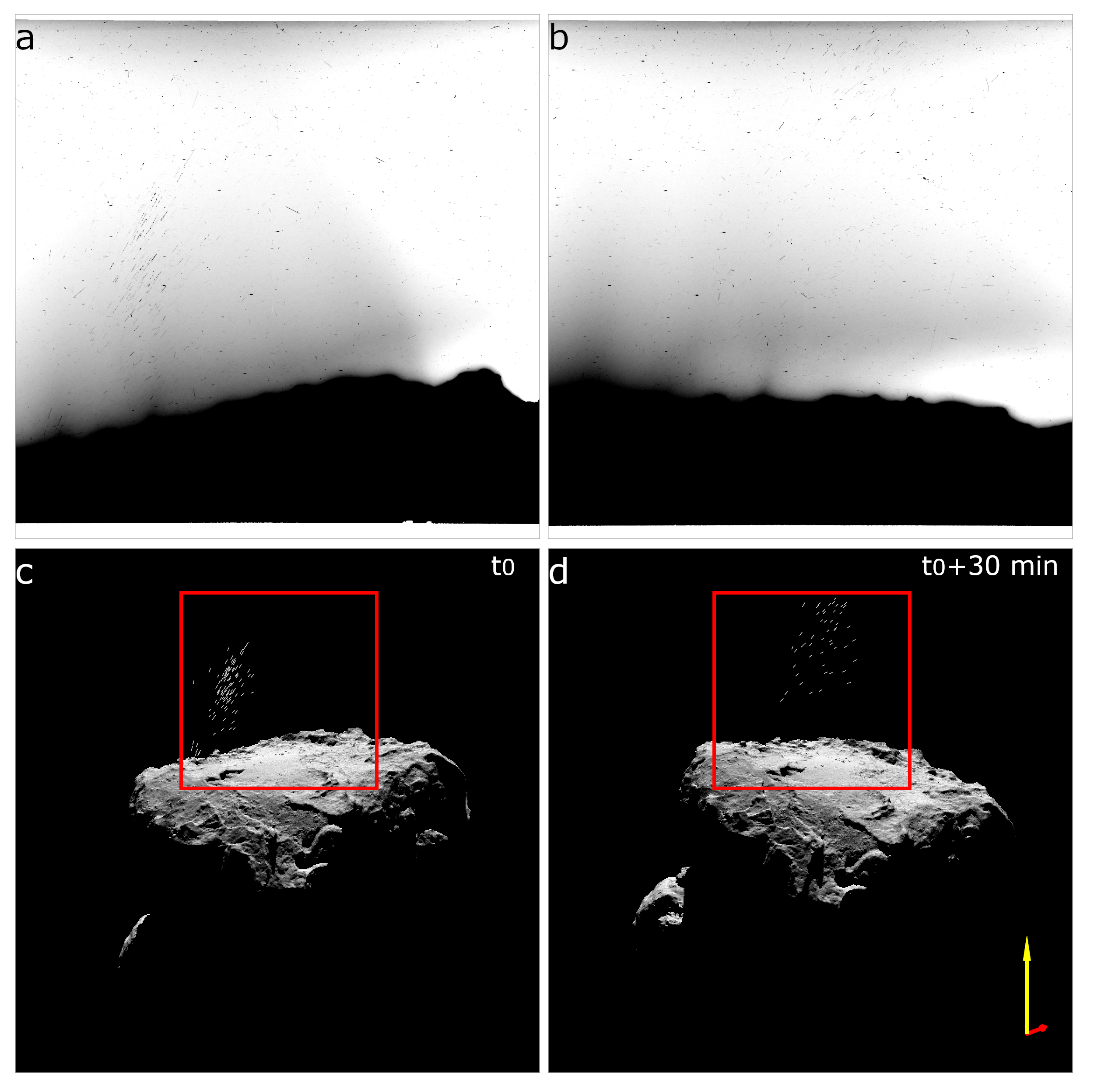} 
\caption{\textbf{Cluster ejection of boulders from 67P observed by OSIRIS on 2016 May 7th.} (a) Image taken by the Wide Angle Camera in reverse grey scale showing clustered tracks produced by ejected boulders from the nucleus surface. (b) Image taken thirty minutes later shows the dispersion of boulder clusters. (c) and (d) are synthetic views showing the camera’s field of view (red rectangles) of (a) and (c) relative to the nucleus, respectively. Identified tracks for photometric analysis are masked as white line segments. Directions of the Sun and of 67P’s rotational axis are indicated by the yellow and red arrows in (d), respectively.}
\label{fig:observation}
\end{figure}

Between March and June 2016, multiple events of clustered ejections of decimeter- to meter-sized boulders were captured by the Wide Angle Camera (WAC) of Rosetta's Optical, Spectroscopic, and Infrared Remote Imaging System (OSIRIS, \cite{2007SSRv..128..433K,2015Sci...347a1044S}) during its ``ballistic" operation mode (Appendix \ref{app:ballistic}). Figure \ref{fig:observation} shows such an observation when the camera was pointing above the Imhotep region on the large lobe of the nucleus\footnote{Please refer to \citet{2015Sci...347a0440T} for the naming and definition of regions on 67P}. In the contrast-enhanced view (Figure \ref{fig:observation}a), a family of streaks is identified in the near-nucleus space, marking the motion of dozens of objects (Figure \ref{fig:observation}c). Another cluster with fewer members is located just above the horizon of the nucleus. Both clusters seem to have originated from the same surface area to which the extensions of the tracks converge. In a following image taken roughly thirty minutes later (Figure \ref{fig:observation}b), both clusters had propagated further away from the nucleus and were dispersed (Figure \ref{fig:observation}d).

\begin{figure}[ht]
\centering
\includegraphics[width=\textwidth]{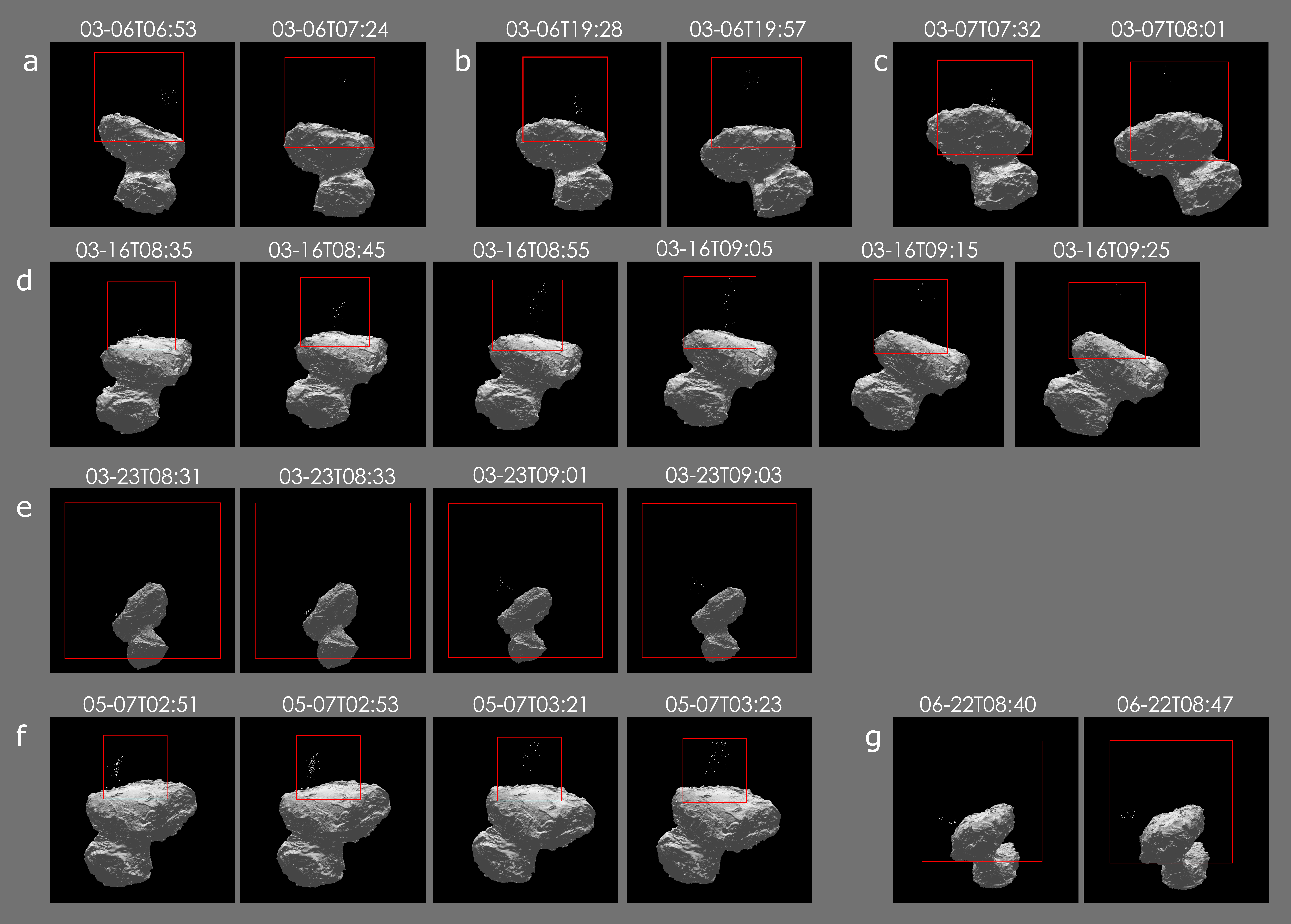} 
\caption{\textbf{Boulder tracks identified in periodic ejection events.} (a)-(c): Sequence on March 6-7, 2016. Three consecutive events were captured at the same rotational phase of the nucleus. (d) Sequence on March 16th, with a relatively high imaging cadence of 10 minutes, covering the expansion of the ejected boulder cluster. (e) Sequence on March 23rd, the southern hemisphere equinox. Rosetta was further away from the nucleus at a distance of $\sim$ 45 km. (f) Sequence on May 7th, as presented in Figure \ref{fig:observation}. (g) The last such event identified on June 22nd.}
\label{fig:observation_all}
\end{figure}

A further search through the OSIRIS data set revealed at least seven such events, all showing clustered bright tracks in the near-nucleus coma (Figure \ref{fig:observation_all}). The earliest event identified occurred on March 6, 2016 (Figure \ref{fig:observation_all}a), and the last one on June 22 (Figure \ref{fig:observation_all}g), covering a range of heliocentric distances from 2.52 AU to 3.26 AU. Particularly, in one sequence covering three rotations of the nucleus, we found repetitive occurrences at almost exactly the same phase of every rotation (Figure \ref{fig:observation_all}a-c). Such a diurnal pattern is usually considered to be the signature of activity driven by water-ice sublimation within the diurnal thermal skin depth of the nucleus.

\section{Analysis} \label{sec:results}
\subsection{Photometry} \label{subsec:photometric}

\begin{figure}[ht]
\centering
\includegraphics[width=0.8\textwidth]{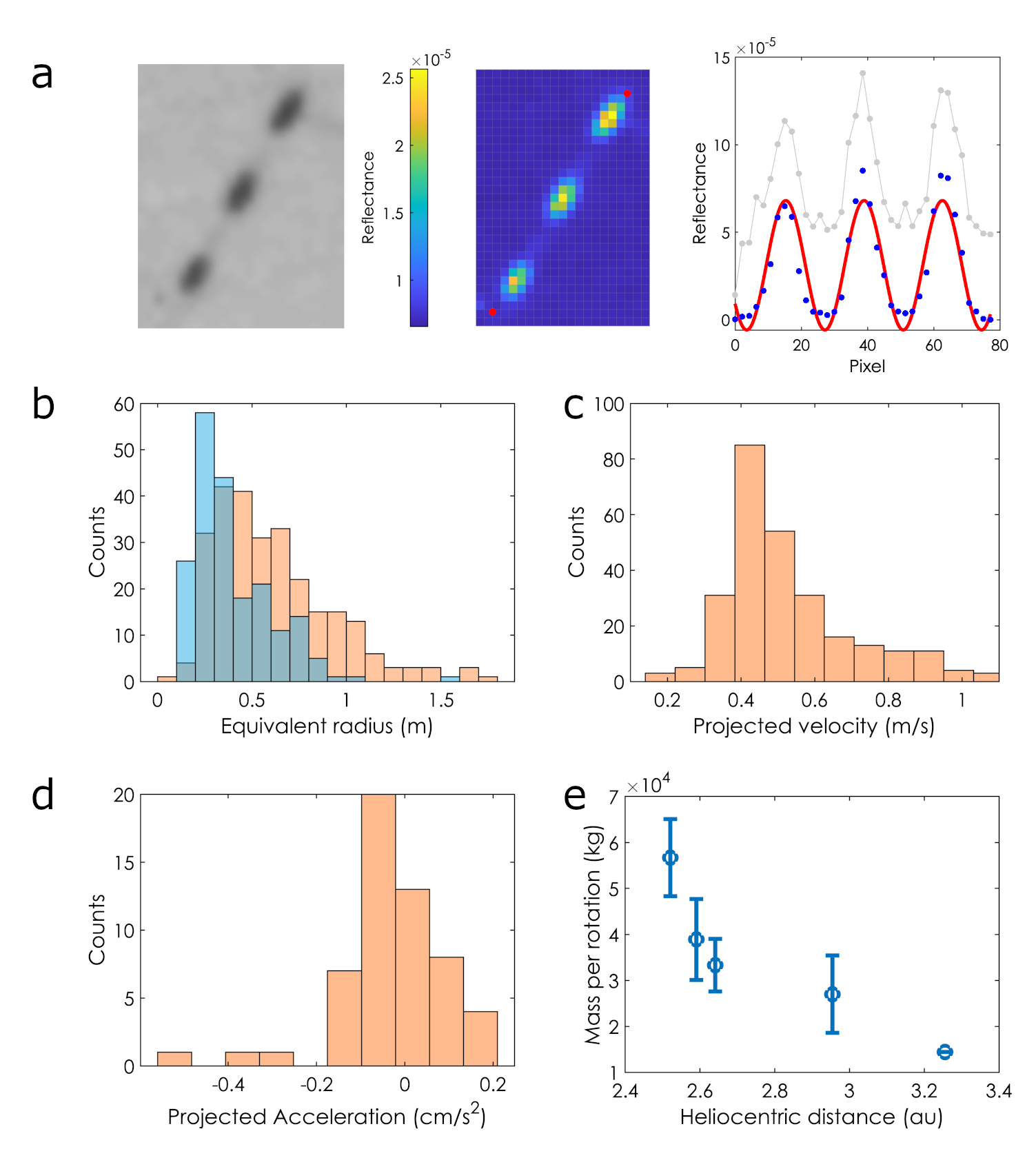} 
\caption{\textbf{Photometric reduction and statistics} (a) Photometric analysis of a typical boulder track. Left: crop of the image (in reverse greyscale) containing the track at centre; middle: pixelated observation with reflectance re-coded in colour; right: extracted light curve along the track, showing the original (grey), the background-removed reflectance (blue dots), and the fitted periodic function (red). The estimated radius is 0.88 m, and the speed 0.46 m/s. (b) Number distribution of the equivalent radii of boulders, assuming that the boulders have homogeneous albedo (orange) or heterogeneous albedo (blue). (c) Number distribution of their projected velocities on the image plane. (d) Number distribution of the projected accelerations of 55 boulders observed in more than one image. (e) Average ejected mass per event as a function of heliocentric distance. Error bar indicates the uncertainty of ejected mass observed during several rotations at the same heliocentric distance.}
\label{fig:statistics}
\end{figure}

Altogether 268 individual boulders were photometrically analysed. Most tracks show periodic variations in brightness (Figure \ref{fig:statistics}a), caused by rotation of the elongated shapes, or varying surface properties, or both \citep{2015A&A...583A..14F}. Assuming that these boulders have the same photometric properties as the average nucleus surface and that they were at the same distances from the spacecraft as the nucleus center, we derived their sizes from the measured reflectance of the tracks (Appendix \ref{app:photometry}). The distribution of equivalent radii peaks at 0.4 m while showing an excess towards the larger end (orange bars in Figure \ref{fig:statistics}b). On the other hand, the brightness variations along the tracks could be at least in part caused by nonuniform surface albedo, e.g., due to exposed ice. In case the maximum brightness had resulted solely from enhanced albedo over a half of the boulder rotation, the peak size would be moderated at $\sim$ 0.3 m (blue bars in Figure \ref{fig:statistics}b). The projected velocities of the boulders are estimated to vary between 0.2-1.1 m/s, peaking at $\sim 0.4 \text{ m/s}$ (Figure \ref{fig:statistics}c). 55 out of the 268 boulders appeared in more than one observation. This enabled us to derive their projected accelerations ranging from -0.5 to 0.2 cm/s$^2$, with a slight majority of the boulders having negative accelerations, i.e., visibly slowed down in the ascent (Figure \ref{fig:statistics}d). The distribution and magnitude of accelerations are similar to those derived from earlier studies in which decimeter-sized boulders were observed to experience deceleration and in one case fall back towards the nucleus \citep{2016MNRAS.462S..78A,2022A&A...659A.171P}. 

If these chunks have the same bulk density as the nucleus of 537.8 kg/m$^3$ \citep{2019MNRAS.483.2337P}, the total ejected mass during one rotation is on average ~40$\times 10^3$ kg, roughly 8\% of the measured global dust production rate of 67P at 2.9 AU \citep{2016A&A...587A.155M}. This rate decreases from 57$\times 10^3$ kg at 2.52 AU to 14$\times 10^3$ kg at 3.26 AU (Figure \ref{fig:statistics}e). The trend indicates that the observed activity of boulder ejections followed steadily the decrease of insolation as the comet withdrew from the Sun, and may represent a regular, seasonal response over the comet orbit.

\subsection{Source and ejection conditions} \label{subsec:source}

\begin{figure}[ht]
\centering
\includegraphics[width=\textwidth]{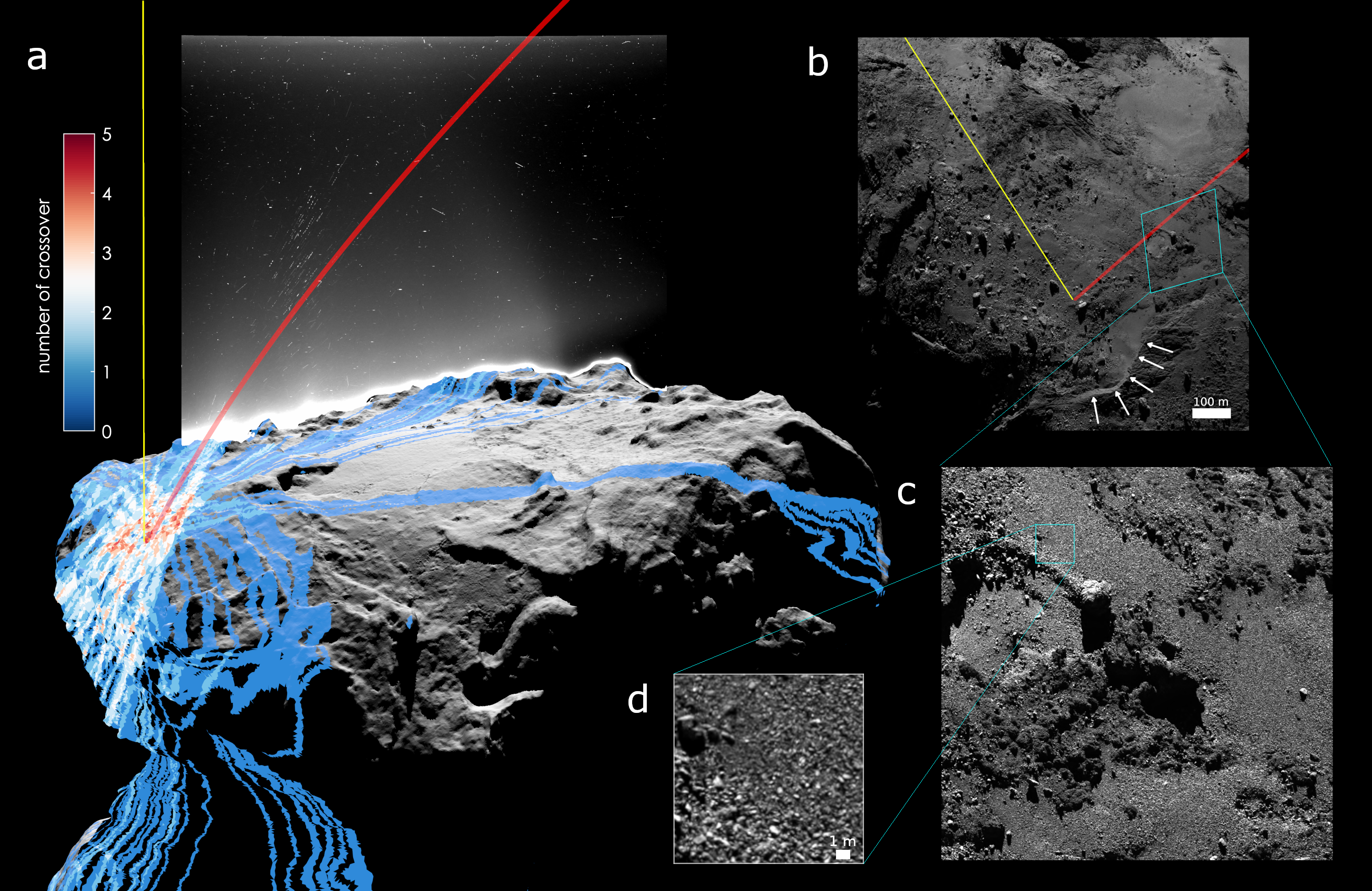} 
\caption{\textbf{Derived common source area of ejected boulders} (a) Synthetic view of 67P over the Imhotep region on the large lobe. Colored stripes over the surface trace potential source locations of the boulders within the planes of their projection onto the images. The color of an area represents the density of count of strip crossovers. The stripes converge in a common area in Bes with the highest concentration of crossovers marked in red. The yellow line indicates the direction of the Sun. The red curve emanating from the probable source region represents a simulated boulder trajectory coincident with the observed tracks in the observation on May 7th, montaged in the background. (b) Bird’s eye view of the source area in Bes. White arrows indicate the edge of a cliff nearby. The red and yellow lines indicate respectively the projected boulder trajectory and solar direction as in (a). (c) High resolution image (9 cm/pixel) of a portion of the source area inside the turquoise quadrilateral in panel (b). (d) Enlarged view of the area in the rectangle in (c) showing in more detail local morphology, a debris field populated with decimeter- to meter-sized boulders.}
\label{fig:source}
\end{figure}

The varying geometries of these observations with respect to the nucleus allow us to probe if there is a common source area. Each observed track and the instantaneous camera boresight direction determine a plane encompassing the boulder motion vector in space. Assuming straight trajectories, the source of the ejection lies along the intersection of the plane with the nucleus and always in the opposite direction of boulder motion (Appendix \ref{app:source}). We find all the resultant intersection profiles converge in one area, roughly 500 $\times$ 500 m$^2$, around (41$^\circ$S, 93$^\circ$E) in the Bes region on the large lobe of the nucleus (Figure \ref{fig:source}a\&b). High resolution images show this area to be scattered with decimeter- to meter-sized rocks of various shapes (Figure \ref{fig:source}c\&d). Moreover, the source area is at the foot of a cliff (indicated by white arrows in Figure \ref{fig:source}b). Cliff collapses, which have been observed at different locations on 67P, are known to replenish the nucleus surface with fresh, icy materials \citep{2017A&A...607L...1P}.

The Bes region has been well noted for comprising relatively high abundance of water ice \citep{2017MNRAS.469S..93F,2017MNRAS.469S.582O}. The source area was also identified as one of the most active areas on the nucleus’ southern hemisphere, manifested in strong regular emissions of H$_2$O and CO$_2$ gas as well as several outbursts \citep{2016MNRAS.462S.184V,2017MNRAS.469S..93F,2017MNRAS.469S.582O,2019MNRAS.483..852L,2016MNRAS.462S.533L}. It is worth noticing that if the ejected boulders contain a fraction of ice twice as much as the average nucleus, their albedo could also be a factor of two higher than the bulk value of the nucleus \citep{2017MNRAS.469S..93F}. In this case, their actual equivalent radius would be $\sim$1.4 times smaller. 

Forward modelling of trajectories from the source area suggests that, to reproduce the observed boulder tracks, their initial velocities must be inclined against the local surface normal. One such typical trajectory is shown in Figure \ref{fig:source}a, whose initial velocity is at a minimum angle of 60 degrees relative to the surface normal. A search through a broader parameter space shows that synthetic trajectories that yield similar tracks as in Figure \ref{fig:observation}a require the initial velocities to be inclined against the local normal by 40 to 70 degrees. This contrasts the persistent dust emissions, such as those in the background, where smaller particles are released perpendicularly from the surface, and follow the bulk gas flow \citep{2018NatAs...2..562S,2018AdPhX...304436K}.

\begin{figure}[ht]
\centering
\includegraphics[width=0.7\textwidth]{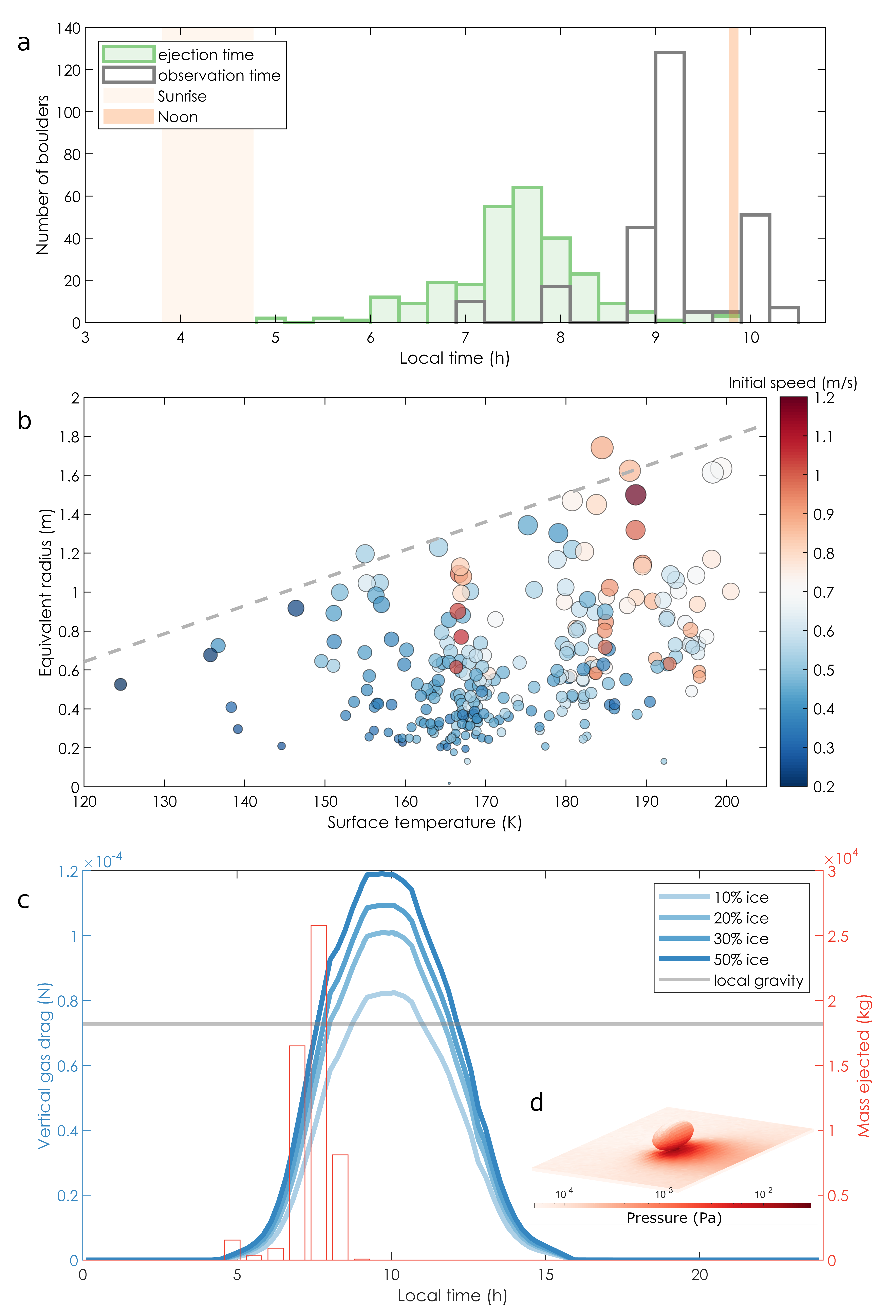} 
\caption{\textbf{Thermal conditions of boulder ejection} (a) The distribution of the observation and inferred ejection times of boulders translated to 24-hour clock with respect to the rotation period of 67P, as indicated by the grey-edged and green bars, respectively. Light orange bar indicates the sunrise times at the source area; the width results from the variation of sunrise times through the time span of observations. Dark orange bar indicates the time when the Sun is at zenith. (b) Relation between the equivalent radius of observed boulders and the corresponding modelled surface temperature at their ejection. Each boulder is represented by a circle, whose size is proportional to the equivalent radius, and whose colour indicates the initial speed. If multiple observations exist for one boulder, the acceleration is estimated and used to infer the boulder’s initial speed; otherwise, the initial speed equals the observed projected velocity. The grey dashed line traces a linear trend of the largest boulder size that could be ejected for a certain surface temperature as inferred from the data. (c) and (d) show ejection scenarios for a 20$\times$8$\times$8 cm ellipsoid-shaped boulder in the source area. (c) Vertical component of gas drag on the boulder over one comet rotation, for different surface ice fractions as indicated by blue curves. The grey horizontal line indicates the magnitude of gravity with centrifugal effect due to nucleus rotation. The bars outlined in red represent the distribution of the inferred total ejected mass against local time during the event on May 7th. (d) Modelled water vapour pressure around the boulder at the time of observation on May 7th.}
\label{fig:ejection}
\end{figure}

We estimated the local time of each boulder’s ejection by assuming it travelled from the inferred source area along a straight line at a constant speed equal to its projected average velocity. Although the seven events spread over three months' time, all of them happened during local morning (Figure \ref{fig:ejection}a). Resulting ejection times show an approximate normal distribution between 5 and 10 am, peaking around 7.30 am (Figure \ref{fig:ejection}a), which coincides with the diurnal heating-up process of the source area. 

We applied a generic thermophysical model with water ice homogeneously distributed in the top layer of the nucleus to reconstruct the thermal condition at the source area (Appendix \ref{app:thermal}). It is found that, at the derived ejection times, the source area had a surface temperature spanning between 120 K and 200 K (Figure \ref{fig:ejection}b). While smaller boulders, with equivalent radius below $\sim$5 dm, were ejected at all temperatures, larger boulders were more likely to have higher inferred ejection temperatures. The maximum size of observed boulders has an approximate linear correlation with the inferred surface temperature at ejection (grey dashed line in Figure \ref{fig:ejection}b). Moreover, boulders ejected at higher temperatures are found to have higher initial speeds. All boulders with initial speed higher than local escape velocity (reddish dots in Figure \ref{fig:ejection}b) were ejected when the source area was warmer than 165 K. Intriguingly, unlike most dust release activities on comets, no clear anti-correlation dependence is found between the size and initial speed of a boulder. One interpretation is that boulders underwent certain accelerating processes after being aloft. Else, they may have left the surface with an initial speed increasing with size.

\subsection{Possible mechanism} \label{subsec:mechanism}

Since the ejections follow closely the thermal condition of the source area, they were most likely part of the regular response of the nucleus to the varying insolation. The actual levitation mechanism could be a combination of multiple driving forces, but the diurnal cyclicity alludes to water ice sublimation being the controlling factor or at least a main trigger. To assess the feasibility for a decimeter-sized boulder to be ejected by gas pressure produced solely from water ice sublimation, we constructed a $20\times8\times8$ cm prolate spheroid-shaped boulder (Figure \ref{fig:ejection}d). It rests on the surface of the source area and has the same homogeneous surface ice content as its surroundings. We modelled its thermophysical state in the realistic insolation environment during the observed event on May 7 2016.

Calculations of the boulder's diurnal illumination condition show that it was subject to a highly asymmetric thermal cycle. The accumulated insolation on the boulder's sun-facing side during one rotation is 4.5 times higher than that on its anti-sun side. In fact, during the three months in which the observations were taken, the maximum elevation angle of the Sun in the source area was constantly below $\sim40^\circ$. Highest temperatures as well as water vapour flux rate are found on the lower part of the boulder’s sun-facing side and its surroundings, leading to asymmetric pressure build-up. We performed Direct Simulation Monte Carlo modelling of the three-dimensional local sublimating gas field, including molecule-molecule collisions as well as interactions with the surrounding surfaces, to assess quantitatively the pressure field around the boulder and in particular inside the narrow crevice between the boulder and the ground (Appendix \ref{app:thermal}). Results show that, at the time of observation, the vapour pressure on the sun-facing side of the boulder is two orders of magnitude higher than that on the opposite side (Figure \ref{fig:ejection}d). With a homogeneous surface ice content of 10\%, the total drag force on the boulder, calculated by integrating the gas pressure over its entire surface, is sufficiently high to overcome gravity (Figure \ref{fig:ejection}c). An increase in iciness would yield stronger forces that could mobilize the boulder at an earlier hour. 
Actually, similar processes, but on a much larger scale, could play a role in 67P's web of surface fractures. They gain higher temperatures and sublimation rates compared to flat terrains \citep{2017A&A...608A.121H}.
It is worth mentioning that, higher ice abundances would result in increased albedo \citep{2017MNRAS.469S..93F}, which could reduce the insolation absorbed by the nucleus. We found that, for 50\% ice, such effect could reduce the water flux rate by $\sim$15\% (Appendix \ref{app:thermal}). However, since the surface of 67P is in general very dark, with an average geometric albedo of less than 6\% \citep{2015A&A...583A..30F}, variation in albedo has a limited influence on ice sublimation rate.

\section{Discussion} \label{sec:discussion}

The water-only ejection scenario could reproduce several important characteristics of the observed events. First, the drag force surpasses local gravity at mid-morning around 7 am, which is consistent with the distribution of the boulder ejection time (Figure \ref{fig:ejection}c). Second, the enhanced gas pressure on the sun-facing side would set the boulder off with an inclined initial velocity, thus reproduce the observed tracks. Moreover, compared to zenith illumination, the slanted illumination subdues the force pressing directly on the boulder caused by ice sublimation from its top, which facilitates its ejection. Boulders under similar thermal condition fulfil the ejection criteria at a similar time and form the clustered launch. However, it might only serve as a limiting case for two reasons: First, the size of the modelled boulder with equivalent radius of $\sim$12 cm, is on the lower end of the observed distribution. Second, the efficiency of pressure build-up between the boulder and its surrounding areas depends on local topography. In cases where such gaps are missing, e.g. with half-buried boulders, the build-up of sufficiently high pressure could prove difficult. 

Overall, water activity alone appears to be insufficient for the ejection of larger boulders and other mechanisms are most likely at work at the same time. Previous works have investigated different types of accelerating mechanism of dust particles around other cometary nuclei \citep{2009Icar..203..571R,2019Icar..325...94V}. In particular, the source area is a confirmed site of strong CO$_2$ emission around the time of the observations \citep{2019MNRAS.483..852L}, whose sublimation flux at a lower temperature of 105 K would already be sufficient to lift boulders of 0.5 m in radius \citep{2009Icar..203..571R,2006hgdc.conf.....H}.  Previous study proposed CO$_2$ sublimation as a regular, repetitive activity of 67P during perihelion, albeit under the drastically different condition during polar day \citep{2020MNRAS.493.3690G}. 

What remains to be understood is how such activity, often sourced from deeper reservoirs, could repeat itself on a diurnal basis without intensive resurfacing events at already large heliocentric distances. One possibility is that it was indeed induced by ice buried at some decimeters beneath the surface and occurred continuously little influenced by diurnal insolation \citep{2022NatAs...6..546C}. This way it provided a persistent background flow assisting boulder lift, whereas the diurnal periodicity was still governed by water sublimation. Another possibility is that CO$_2$ sublimation itself varied diurnally as water activity. Because clear evidence of its diurnal behaviour is lacking, local CO$_2$ activity as a background lifting force that varied seasonally seems to be more intuitive. Nevertheless, the possibility of diurnal CO$_2$ sublimation should by no means be precluded but deserves careful investigations via rigorous thermophysical modelling. 

Acceleration could also be provided by the sublimation of volatile ices contained within the boulder itself \citep{2009Icar..203..571R}. This so-called ``rocket force" was already discovered on decimeter-sized chunks in the inner coma of 67P \citep{2016MNRAS.462S..78A}. It is conceivable that, given the irregular shape, the bottom part of a boulder retains considerable amount of frost from night-time deposition. Once the boulder is mobilized and the frost is exposed to illumination, the rocket force is generated.

The local topography could play a decisive role in controlling the ejection direction of a boulder. The setting of our model, i.e. an ellipsoidal boulder sitting on a flat surface, is much simplified compared to realistic boulders and their environments on 67P. As shown in Fig. \ref{fig:source}d, the boulder field at the source area consists of decimeter- to meter-sized boulders of all kinds of shapes and are embedded in their surroundings in different ways. Further studies are needed to reconstruct these ejections with realistic setting for unveiling details of such boulder activity.

Rather than being responsible for the comet's mass loss, such activity would contribute primarily to the global material recycling and relocation on the nucleus \citep{2015A&A...583A..17T,2017A&A...604A.114H,2017MNRAS.469S.357K}. The source area is dynamically flat, i.e., close to be gravitationally equipotential, with an escape velocity between $0.6-0.8 \text{m/s}$ (Appendix \ref{app:source}). This suggests that half of the ejected boulders would ultimately fall back on the nucleus, providing a mechanism for boulders and extant volatile ices therein to be transported from the foot of a cliff to an extended area on the nucleus. However, since a specific illumination geometry is required for such ejection to happen, different regions would undergo such kind of activity when 67P was at different positions on its orbit.

We anticipate this type of activity to occur on a range of weakly gravitated icy bodies. It could happen either around perihelion for bodies with low ice content and up to large heliocentric distances for ice-rich bodies. As long as a favourable illumination geometry is reached, chunks could launch from the surface, forming perceivable mass-shedding phenomena.
\bigbreak
\textbf{Acknowledgements}  J.A. acknowledges funding from the European Union's Horizon 2020 research and innovation programme under grant agreement No 757390 CAstRA and from the Volkswagen Foundation. OSIRIS was built by a consortium led by the Max-Planck-Institut f{\"u}r Sonnensystemforschung, G{\"o}ttingen, Germany, in collaboration with CISAS, University of Padova, Italy, the Laboratoire d'Astrophysique de Marseille, France, the Instituto de Astrof{\'i}sica de Andalucia, CSIC, Granada, Spain, the Scientific Support Office of the European Space Agency, Noordwijk, The Netherlands, the Instituto Nacional de T{\'e}cnica Aeroespacial, Madrid, Spain, the Universidad Polit{\'e}chnica de Madrid, Spain, the Department of Physics and Astronomy of Uppsala University, Sweden, and the Institut f{\"u}r Datentechnik und Kommunikationsnetze der Technischen Universit{\"a}t Braunschweig, Germany. The support of the national funding agencies of Germany (DLR), France (CNES), Italy (ASI), Spain (MEC), Sweden (SNSB), and the ESA Technical Directorate is gratefully acknowledged.

\appendix

\section{``Ballistic stacked" observations by OSIRIS WAC camera}\label{app:ballistic}
The shutter of the OSIRIS Wide Angle Camera was operated in the special ``ballistic" mode following the breakdown of its locking mechanism. Observations used in this study were acquired based on this mode, acquired with the so-called ``ballistic stacked" technique. Basically, the shutter was commanded to perform a certain number of ballistic operations, resulting in a series of short exposures of $\sim$150 ms with $\sim$1.5 s in between. The image retrieved was thus the stack of these short exposures. For example, Figure \ref{fig:observation}a is a stack of 50 ballistic exposures. Its equivalent exposure time is 7.5 s, while the observation actually lasted for over 77 s. WAC images taken with a ballistic mode of its shutter were calibrated in a special manner. Detailed description could be found in the calibration documents accompanying the published OSIRIS dataset. We noticed that the first event identified was recorded within less than a month after the ballistic mode was first commanded. The phenomena thus could have well preceded their observations that were more easily or directly captured with the new mode. 

\section{Photometric analysis of boulder tracks} \label{app:photometry}
We used OSIRIS level 3 (CODMAC level 4) data of calibrated reflectance images for the photometric analysis. The images were geometrically and radiometrically calibrated. Details of the calibration pipeline are described in \citet{2015A&A...583A..46T}. Boulder tracks were identified and catalogued via visual inspections. In the crop-out of the image containing each track, we extracted the light curve of the moving boulder along the line defined by the visually identified starting and end points. Most curves show clear periodic variations in brightness due to an irregular shape or heterogeneous surface properties. This variation in brightness is used to estimate the boulder’s rotational state by fitting a periodic function. If we picture the boulder as an ellipsoid, its spin rate is half the frequency of the periodic function. The projected velocity on the image plane was estimated as the ratio between the length of the track and the time span of the observation.

To estimate sizes of the boulders, we applied aperture photometry with the method developed by \citet{2017MNRAS.469S.312G}. Accumulated reflectance of the boulder is calculated by integrating signal within the smallest aperture enclosing the identified track. The local background level was estimated as the average signal level in the annulus immediately outside the aperture. The noise-subtracted reflectance of the entire track was then used for estimating the boulder’s dimension. By assuming the boulder has the same photometric property as the integrated nucleus surface, and, by assuming that the boulder locates at the same distance as the nucleus center relative to the spacecraft, the equivalent radius of the boulder $r$ is calculated as \citep{2017MNRAS.469S.312G}:

\begin{equation}
    r = \sqrt{(\frac{1}{\pi}\frac{R}{R_0})}\cdot \frac{s_P}{f}\cdot d 
\end{equation}
\begin{equation}
    R_0 = 0.055\cdot \exp{-0.045\cdot \phi}
\end{equation}

Here $R$ is the boulder's reflectance, $s_P=13.6 \mu m$ is the pixel size of WAC's detector, $f = 135.68$ mm is the focal length of WAC, $d$ is camera's distance to nucleus' center in meter. $R_0$ is the nucleus' reflectance calculated with the phase angle $\phi$ at the time of the observation. 
 
\section{Determination of the source area} \label{app:source}
Assuming that the boulders move along a straight line in the near-nucleus space, its trajectory would fall in a plane defined by the position of the camera and the boulder's projected track on the image plane. By intercepting this plane with the nucleus, we could infer the possible source point of the boulder which should be a certain point located on the intersection line. Similar methods have been widely adopted in studies on the inversion of the sources of cometary jets \citep{2013Icar..222..540F,2016A&A...587A..14V}.

We constructed the trajectory planes of all identified boulders based on their tracks and the camera's position from Rosetta SPICE kernel dataset (\url{https://doi.org/10.5270/esa-tyidsbu}). Then we intercept these planes with a shape model that represents the nucleus with $\sim 500,000$ facets (resolution: $\sim$10 m, \cite{2017A&A...607L...1P}). Since each event was imaged with a different observation geometry, resulting interception lines lie in all directions (Figure \ref{fig:source}a). A common source region was identified where dense intersection points of these lines exist. 

Trajectories of boulders initiating from the determined source region, as shown in Figure \ref{fig:source}a, are modelled assuming the boulder is not coupled with local gas field. We propagate the position and velocity vectors of the modelled boulder by applying gravitational force, centrifugal force, and Coriolis forces at each step. A standard Runge-Kutta (4,5) integrator is used. Gravitational forces are calculated taking into consideration topography of the nucleus. We applied the polyhedral method developed by \citet{1996CeMDA..65..313W} to a 67P shape model with 1000 facets \citep{2017A&A...607L...1P}.

The local escape speed at the source region was estimated according to \citet{1996Icar..121...67S}, considered as the minimum speed a particle needs to evade falling back to the surface. For a point at coordinates $\boldsymbol{r}$ on the nucleus surface, the escape speed $v$ is calculated as \citep{1996Icar..121...67S}:

\begin{equation}
    v = -\boldsymbol{n} \cdot (\boldsymbol{\Omega} \times \boldsymbol{r}) + \sqrt{[\boldsymbol{n}\cdot(\boldsymbol{\Omega} \times \boldsymbol{r})]^2 + 2U - (\boldsymbol{\Omega} \times \boldsymbol{r})^2}
\end{equation}

here $\boldsymbol{n}$ is the unit vector of local surface normal, $\boldsymbol{\Omega}$ is 67P's angular velocity,  $U$ is the surface gravitational potential including centrifugal potential, calculated using the 500,000-facet polyhedron shape model \citep{2017A&A...607L...1P}.

\section{Thermophysical modelling} \label{app:thermal}
A generic one-dimensional thermophysical model was used for synthesizing the thermal condition of both the source area and the modelled spheroidal boulder. Details of the  numerical treatments can be found in \citet{2017MNRAS.469S.295H,2019A&A...630A...5H,2021ApJ...910...10H}. 

For the source area, we used a regional shape model with $\sim$20,000 facets that represents the Imhotep region and its surroundings with a resolution of $\sim$30 m. The model is able to account for shadowing and self-heating effects caused by neighbouring topographic features. We modelled the thermal condition of the area at the time of each ejection event by simulating the diurnal cycle of water ice activity at a time step of $\sim$300 s. The ephemerides of 67P and its rotation state were retrieved from the above-mentioned Rosetta SPICE dataset. A homogeneous ice content of 10\% is assumed in accordance with the relatively high abundance of water ice in this area found by previous studies. 

For the boulder, we first constructed a polyhedral shape model consisting of a spheroid dwelling on a surrounding surface plane. This local setting is then placed in the actual environment of the source area via a coordinate system transformation. To set up this scenario as a limiting case, we rotate the boulder azimuthally so that its long axis is perpendicular to the projected direction of the Sun at local midday, hence maximizing its cross section for energy absorption. Surface temperature as well as water ice outgassing rate were modelled in the same fashion as thermal modelling of the source region. They were applied as boundary condition for the gas field modelling described below.

To assess the thermophysical impact of albedo variation, we modelled diurnal water sublimation rate of a representative facet on the boulder with both constant and iciness-dependent albedo. The iciness-dependent albedo $A$ is calculated assuming a simple linear relationship with ice abundance, so that $A=0.012$ corresponds to 10\% of water ice. 
Model results confirm that a higher albedo indeed reduces the absorbed insolation energy, and consequently reduces the water flux rate. For the case of 50\% ice, the maximum water flux with an iciness-dependent Bond albedo of 0.06 is ~15\% lower than that modelled with a constant albedo of 0.012. This effect gets weaker with lower ice abundances. For the case of 20\% ice, the difference is roughly 5\%. On the other hand, varying albedo based on iciness does not change the overall trend of flux rate, that is, higher abundance of ice results in higher water vapor flux rate, which we believe is mainly due to the very low reflectance of 67P's surface. Also, even reduced due to higher albedo, the water flux rate still exceeds that produced by 10\% ice, meaning that the total gas drag force should still be sufficient for mobilizing the boulder.

To assess the pressure exerted on the boulder by the vapour from water ice sublimation, we simulated the diurnally varying three-dimensional gas field surrounding the boulder using the method of Direct Simulation Monte Carlo (DSMC, \cite{1994mgdd.book.....B}). For a certain time point, we assume each facet releases gas at a steady rate as derived from the above thermophysical modelling. Initial temperature of the gas is set to be the same as the surface temperature at its source. The DSMC method simulates the motion of gas molecules kinetically. We thus estimate the gas drag force on each facet of the boulder by calculating momentum flow of gas molecules impinging on it. The overall gas drag force exerted on the boulder was estimated as the summation of drag force vectors of all facets.

\bibliography{ApJL.2023.Shi}{}
\bibliographystyle{aasjournal}

\end{document}